\documentclass[
  twocolumn,
  prb,
  nopacs,
  amsmath,
  amssymb,
  superscriptaddress
]{revtex4}
\usepackage{color}

\usepackage{placeins}
\usepackage{hyperref}
\usepackage{amsfonts}

\usepackage{amsmath}
\usepackage{amssymb}
\usepackage{graphicx}
\usepackage{empheq}
\usepackage{subfigure}
\usepackage[usenames,dvipsnames]{xcolor}
\usepackage{soul}
\usepackage[title]{appendix}

\renewcommand{\Re}{\mathop{\mathrm{Re}}}

\newcommand{\beq}{\begin{equation}}
\newcommand{\eeq}{\end{equation}}
\newcommand{\beqa}{\begin{eqnarray}}
\newcommand{\eeqa}{\end{eqnarray}}
\newcommand{\Beqa}{\begin{eqnarray*}}
\newcommand{\Eeqa}{\end{eqnarray*}}
\newcommand{\nn}{\nonumber}

\begin{document}
\title{Thermal Transport in One Dimensional Electronic Fluid }
\author{R. Samanta}
 \affiliation{Department of Physics, Bar Ilan University, Ramat Gan 52900, Israel}
 \author{I.~V. Protopopov}
\affiliation{Department of Theoretical Physics, University of Geneva, 1211 Geneva, Switzerland  }
\affiliation{Landau Institute for Theoretical Physics, 119334 Moscow, Russia}
\author{A.~D. Mirlin}
\affiliation{Institut f\"ur Nanotechnologie, Karlsruhe Institute of Technology, 76021 Karlsruhe, Germany}
\affiliation{Institut f\"ur Theorie der Kondensierten Materie, Karlsruhe Institute of Technology, 76128 Karlsruhe, Germany} 
\affiliation{Petersburg Nuclear Physics Institute, 188350 St. Petersburg, Russia}
\affiliation{Landau Institute for Theoretical Physics, 119334 Moscow, Russia}
\author{D.B. Gutman}
 \affiliation{Department of Physics, Bar Ilan University, Ramat Gan 52900, Israel}
\begin{abstract}
We study thermal conductivity for one-dimensional electronic fluid. The many-body Hilbert space is partitioned into bosonic and fermionic sectors that  carry the thermal current in parallel.  
For times shorter than bosonic Umklapp time, the momentum of Bose and Fermi components are separately conserved,   giving rise to the ballistic  heat propagation and imaginary heat conductivity proportional to $T / i\omega$. 
The real part of thermal conductivity is controlled by decay processes of fermionic and bosonic excitations, leading to several regimes in frequency dependence. 
At lowest frequencies or longest length scales, the thermal transport is dominated by L{\'e}vy flights of low-momentum bosons  that lead to  a fractional  scaling,  $\omega^{-\frac{1}{3}}$  and $L^{1/3}$, of heat conductivity  with the frequency $\omega$ and system size $L$ respectively.

\end{abstract}
\maketitle

In interacting systems, Wiedemann-Franz law is violated and  there is no universal relation between 
the thermal and electric conductivities.
As a result, thermal transport reveals information that can not be accessed by measuring charge transport.
Due to experimental and theoretical challenges,  thermal transport is far less explored compared to charge transport.

In recent years,  the situation started to change, and energy transport
was measured in several experiments. The universal value of thermal conductance $g_0= \pi^2 T/3h $ was observed  \cite{Schwab,Meschke,Jezouin,Cui}  in various devices with ideal point contacts. Heat Coulomb blockade was observed in \cite{Pierre}, directly demonstrating a controllable  energy-charge separation. The propagation of heat in the quantum-Hall-effect regime was intensively  investigated\cite{Yacoby2012,Altimiras2012,Grivnin2014,heiblum1, heiblum2, heiblum3}. Frequency dependence of the thermal conductivity was studied in various materials \cite{koh2007,min2011,regner2013}.

The thermal transport in low-dimensional classical fluids and 1D anharmonic chains (of which the Fermi-Pasta-Ulam-Tsingou model \cite{fpu} is a prominent example)   has been studied in the framework of (non-linear) fluctuating hydrodynamics. 
The corresponding renormalization group (RG) \cite{nrs}  or self-consistent mode-coupling \cite{Spohn2014} analysis predicts  that  the thermal DC conductivity scales with the system size as $\sigma \sim L^{1/3}$ in 1D  models with momentum conservation. This prediction was verified by means of numerical simulations in Ref.  \cite{PhysRevLett.98.184301}. 
The $\sigma\sim L^{1/3}$ behavior of the heat conductivity is closely related to the anomalous broadening, $\Delta x \sim t^{1/z}\sim t^{ 2/3}$, (on top of ballistic propagation) of the sound  peak in the density-density correlation function of the system. The dynamical exponent $z=3/2$ entering here is a manifestation of the  Kardar-Parisi-Zhang (KPZ) fixed point in the RG flow that governs the propagation of the sound mode~\cite{Beijeren2012}. The KPZ scaling of the density-density correlation function was also predicted (within the classical Gross-Pitaevskii formalism) to occur in 1D Bose gas at finite temperature \cite{klm}.  In the context of 1D Fermi systems the KPZ scaling was advocated in Ref.\cite{bovogangardt}.

The experimental progress combined with open fundamental questions prompts us to study  thermal transport in a 1D  quantum electronic fluid.  
In the low-energy limit such fluids are often described within the Tomonaga-Luttinger model  that linearizes the spectrum of fermions near Fermi points and treats the interaction of fermions as point-like. Via the bosonization procedure\cite{giamarchi}, Tomonaga-Luttinger model  maps  to free bosons with a linear dispersion relation. The Luttinger liquid (LL) fixed point is thus a free theory with essentially trivial kinetics. It turns out however, that the corrections to it, a finite curvature of fermionic spectrum and  a finite range of  interactions, albeit irrelevant in RG sense, can have profound effect on the low-energy behaviour of the system's \emph{dynamical}  correlation functions \cite{imambekov2009,imambekov11,pgom2014, pgsm2013,Khodas2007,fili2016}.
Therefore, those perturbations are to be taken into account in the discussion of the thermal conductivity, i.e., the conventional LL  paradigm is not sufficient for this problem.

With the aforementioned corrections included, the Tomonaga-Luttinger model turns into an interacting theory both in  fermionic and bosonic \cite{Landau, Landau1, schick68, Sakita, Jevicki_Sakita} languages. One then has to resort to perturbative treatment   of the model. 
Two choices  of basis for such a perturbation theory are available\cite{bfduality}: i) one can choose to work with the bosonized version of the theory treating the nonlinearity of fermionic spectrum (that translates into interaction of bosons) perturbatively; ii) the bosonic theory can be refermionized \cite{Rozhkov, imambekov11, bfduality} giving rise to the description of the system in terms of fermionic quasiparticles that  are related to original electrons via non-linear unitary
transformation. In the later approach the curvature of the bosonic spectrum  translates into the interaction of fermions
\cite{footnote}.

It was shown in Ref. \cite{bfduality} that the applicability of perturbation theory for {\it thermal} (with energy of order T) excitations in the fermionic and bosonic approaches depends on temperature. Specifically, the effective mass of fermionic excitations $m_*$  and a length $l$ quantifying the  curvature of bosonic spectrum [see Eq. (\ref{uq})] define a temperature scale $T_{\rm FB}=1/m_*l^2$. At $T<T_{\rm FB}$ the thermal fermions are long-living excitations; the perturbation theory in their interaction is well-behaved and controlled by the small parameter $T/T_{\rm FB}$.  At higher temperatures, $T>T_{\rm FB}$,  the proper thermal excitations are bosons and the bosonic perturbation theory possess  a small parameter $T_{\rm FB}/T$.

In this work we employ the  combination of the fermionic and bosonic frameworks to study low-temperature thermal conductivity of the electronic fluid. It turns out that {\it subthermal} excitations dominate the thermal transport at low frequencies and one is faced with the problem of understanding their kinetics.  We  show that at lowest frequencies the behavior of thermal conductivity is anomalous and  has the universal scaling  
$\sigma(\omega)\propto\omega^{-\frac{1}{3}}$. This corresponds to length dependent DC thermal conductivity $\sigma(L)\propto L^{\frac{1}{3}}$ consistent with the classical hydrodynamic limit\cite{nrs,Spohn2014}. At higher frequencies, we identify a variety of new regimes characterized by  power-law   dependence of thermal conductance on frequency,  temperature and system size.

We consider a model  of spinless right- and left-moving fermions 
\beqa
 H=\sum_{\eta}\int dx \psi^\dagger_\eta(x)\left(-i\eta v_F\partial_x-\frac{1}{2m}\partial_x^2\right)\psi_\eta(x)\nn\\
 +\frac{1}{2}\int dx dx' g(x-x')\rho(x)\rho(x')\,,
 \label{sec2:Hamiltonian}
\eeqa
where $g(x) $ is short-range interaction potential, and the total density $\rho(x)$  is a sum of the chiral components,
$\rho(x)=\rho_{R}(x)+\rho_{L}(x)$. 
In the low momentum limit ($ql\ll 1$), the interaction potential is 
 $g_{q}-g_0\propto q^2l^2$. 
  
After  bosonization, the original  Hamiltonian(\ref{sec2:Hamiltonian})  is mapped to an  interacting bosonic model \cite{hald,stone,delft98,gogolin,giamarchi,bfduality}, see Appendix \ref{bsnz}. 
The interaction between electrons Eq.(\ref{sec2:Hamiltonian})  controls the dispersion  of the bosonic modes. At small momenta
\begin{equation}
\omega_q^B= u_q^B |q|\,, \qquad u_q^B=u \left(1-l^2 q^2\right),
\label{uq}
\end{equation}
where $u=v_F \sqrt{1+g_0/{\pi v_F}}$ denotes the sound velocity.  

We now construct the kinetic equation for bosonic and fermionic quasiparticles.
 These equations can be derived from the fermionic (\ref{sec2:Hamiltonian})  and bosonized  versions 
 of the Hamiltonian in a standard way, see Appendix \ref{ktblin} and Appendix \ref{ktf},
\beqa
\frac{\partial N_{\alpha}(q) }{\partial t} +u^{\alpha}_q \frac{\partial N_{\alpha}(q)}{\partial x}= I_{\alpha,q} [N_\alpha]\,.
\label{kinEq}
\eeqa
Here $\alpha=F/B$ specifies the type of the quasi-particles (Fermi/Bose),
 $N_\alpha$  is a   distribution function and $I_\alpha$ is the collision integral.  
A combination of two equations (\ref{kinEq}) permits to extend the Bose-Fermi duality framework \cite{bfduality} away 
from thermal equilibrium.
We next linearize the kinetic equation using the ansatz 
$N_\alpha = n_\alpha + \delta N_\alpha$, where $n_\alpha$ is the quasi-equilibrium distribution 
and $\delta N_\alpha$ is a deviation from a local equilibrium, see Appendix \ref{ktblin} (for bosons) and Appendix \ref{ktf} (for fermions).
To determine the heat conductivity the linearized kinetic equation should be solved for $\delta N_\alpha$ with the temperature gradient introduced into $n_\alpha$. 
In either fermionic or bosonic approach the energy current  can then be computed as 
\begin{equation}
J_\alpha(\omega)= \int (dq )~u^\alpha_q \omega_q^{\alpha} \delta N_\alpha(q)
\end{equation}
where for fermions 
\begin{equation}
\omega_q^F=\pm u q +q^2/2m^*\,, \qquad u_q^F=\pm u+q/m^*.
\label{omegaF}
\end{equation}
In Eq. (\ref{omegaF}) the $\pm$ sign refers to the right and left movers;  $(dq)\equiv \frac{dq}{2\pi\hbar}$ and we set $\hbar=1$ through the  manuscript.  The explicit formula for the effective mass $m_*$ is given in  Appendix \ref{bsnz}.

Before discussing the relaxation of fermionic and bosonic modes in more detail, we note that the model 
(\ref{sec2:Hamiltonian}) as stated  preserves, apart from the charge conservation, also the {\it difference} of the number of right- and left-moving fermions. In the bosonic language this corresponds to the conservation of the total momentum of the bosonic excitations \cite{matprb12,mick2010,Mat2012,mat2014}. Correspondingly, the linearized collision integral 
(both in the fermionic and bosonic formalisms) has a zero mode that gives rise to a ballistic transport of heat\cite{MatveevAndreev2018}. In a more accurate description of the electronic fluid  the  chiral  branches merge at the bottom of the energy band enabling the equilibration of the number of left and right fermions. Within the bosonic  description  this process  corresponds to the Umklapp scattering. The associated time scale is exponentially long \cite{matprb12,mick2010,Mat2012,mat2014}  

\begin{equation}
\tau_{\rm U}^{-1}\sim T^{3/2} \epsilon_F^{-1/2}e^{-\frac{\epsilon_F}{T}}\,.
\end{equation}
Here $\epsilon_F\sim m v_F^2$ is an  ultraviolet energy scale and we omitted a non-universal numerical coefficient  that is determined  by  interaction strength and by details of the spectrum at  the bottom of the band. 
The contribution of the (almost) zero mode associated to the conservation of the bosonic momentum can then be extracted either in fermionic or bosonic framework \cite{kane96,lev2010}
\beqa
&\sigma^{\rm bal}(\omega)= \frac{\pi}{3}\frac{uT }{i \omega+\tau^{-1}_U} .
\label{bal}
\eeqa
At $\omega \tau_{\rm U}\gg 1$,  $\sigma^{\rm bal}(\omega)$ is purely imaginary and does not contribute to the dissipative real part of the total thermal conductance.  In the opposite limit, $\omega \tau_{\rm U}\ll 1$ the contribution of $\sigma^{\rm bal}(\omega)$ becomes a (large) frequency-independent constant, 
$\sigma^{\rm bal}=\pi \tau_{\rm U}uT/3$. 

Let us now turn to the analysis of the relaxing modes in the kinetic equation (\ref{kinEq}).  Employing the relaxation-time  approximation for its solution we find
\beqa
&\Re\sigma^B(\omega)\simeq \frac{T^4 l^4}{u^2} {\rm Re}\int_0^{T/u} ~   \frac{(d q)}{{\tau^{-1}_{B}(q)} -i \omega },  
\label{maineqb}
\eeqa
for the bosonic and 
\beqa
&\Re \sigma^F(\omega)\simeq\frac{T^2}{m_{*}^2u^2} \Re \int_0^{T/u} ~   \frac{(d q)}{{\tau^{-1}_{F}(q)} -i \omega },
\label{maineqf}
\eeqa
for the fermionic representation of the theory, respectively, 
see  Appendix \ref{btc} and Appendix \ref{frtc}. In Eqs. (\ref{maineqb}) and (\ref{maineqf}), $\tau_B(q)$ and $\tau_F(q)$ denote the relaxation times for the bosons and fermions.   Note, that  the prefactors in Eqs. (8) and (9) match at $T=T_{\rm FB}$, which is a manifestation of 
Bose-Fermi duality \cite{bfduality}.  One thus might think that for $T > T_{\rm FB}$ only bosonic excitations are relevant, and for  $T < T_{\rm FB}$ only fermionic ones.
However, this is not true. As we discuss below, the  bosonic lifetime $\tau_B(q)$ diverges in low q limit, while $\tau_F(q)$ remains constant. This makes the two channels of heat transport profoundly different: the bosonic one {\it always} dominates the low-frequency thermal conductivity, irrespectively of the relation between $T$ and $T_{\rm FB}$ .

Equations  (\ref{maineqb}) and (\ref{maineqf}) represent contributions of bosonic and fermionic quasiparticles to the thermal conductivity. Thus, taking into account also the ballistic contribution $\sigma^{\rm bal}$ discussed above, we approximate the total thermal conductivity of the electronic fluid by
\begin{eqnarray}
\!\! \!\! \sigma(\omega)&=&\sigma^{\rm bal}(\omega)+\sigma^\prime(\omega),\\
\!\! \!\!   \sigma^\prime(\omega)&=&\sigma^F(\omega)+\sigma^B(\omega) \simeq\max\left[\sigma^F(\omega), \sigma^B(\omega) \right] \! .
\end{eqnarray}

\begin{figure}
    \begin{tabular}{cc}
        \includegraphics[width=4cm]{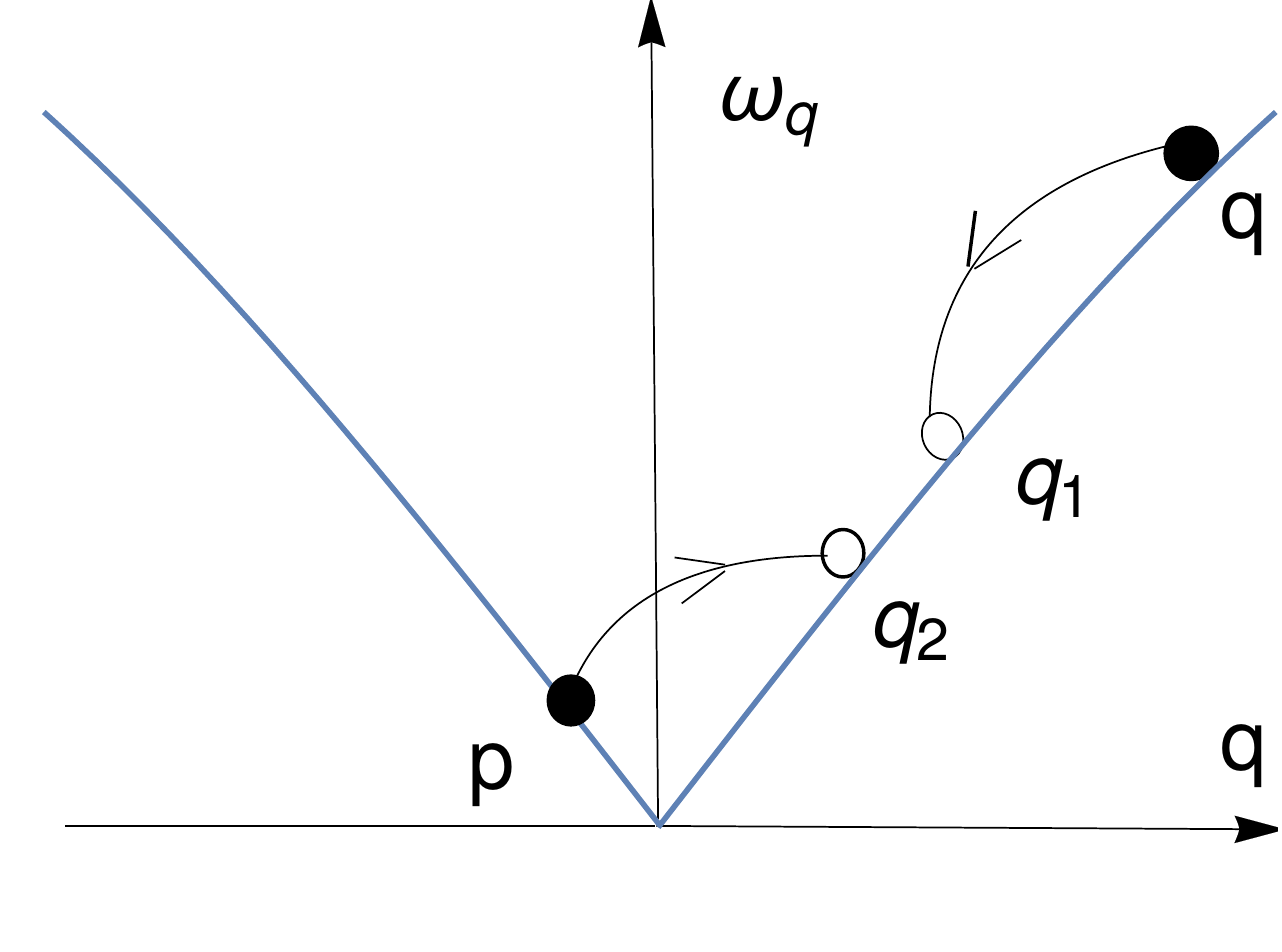}&\includegraphics[width=4cm]{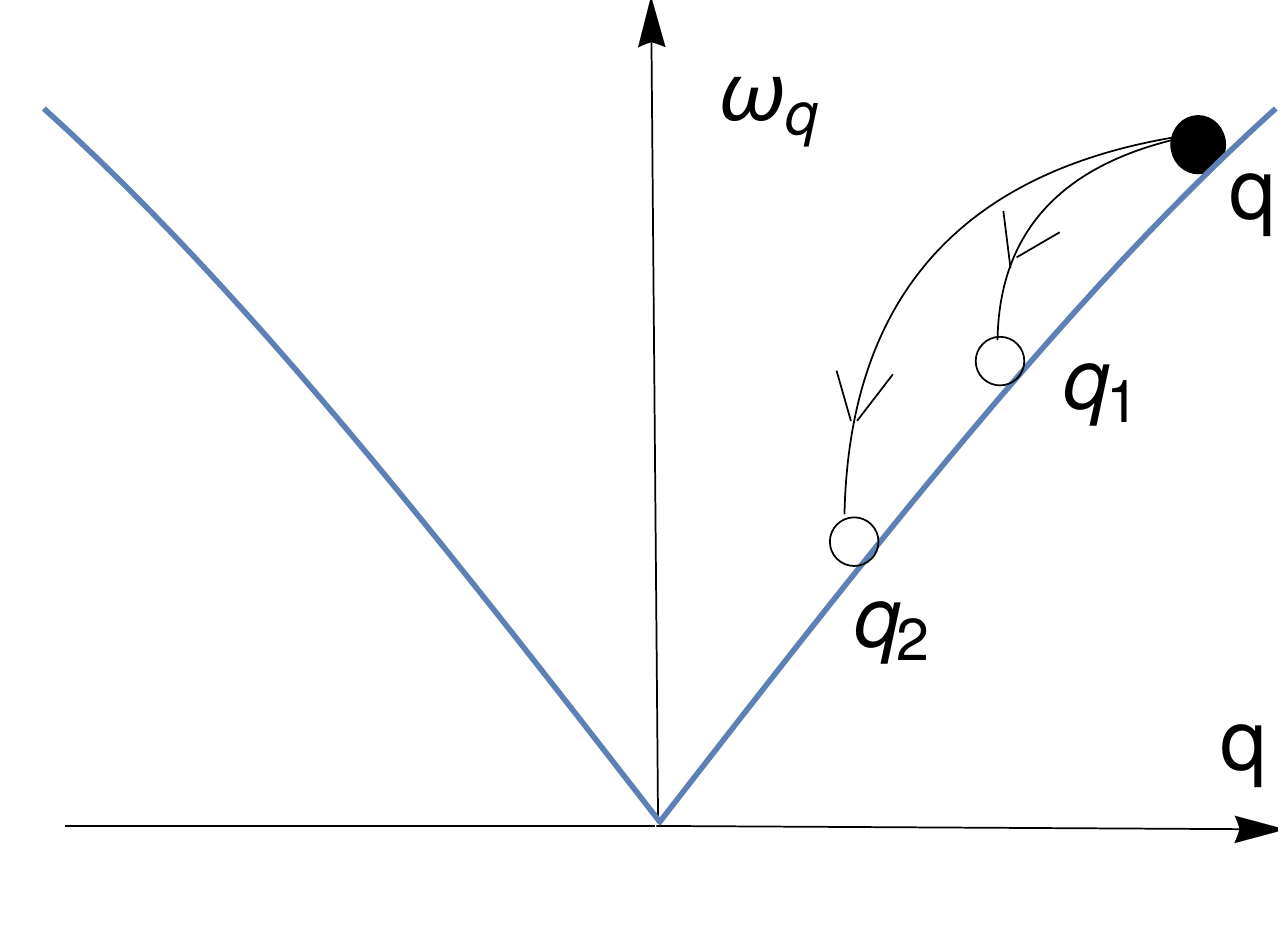}\\
    \end{tabular}
    \caption{
Left: Two-into-two bosonic scattering process involving three bosons of the same chirality  and one boson of opposite chirality.
Right: The self-consistent decay process of one boson into two, yielding
  the decay rate $\tau^{-1}(q)\propto q^{3/2}$ for $q<q_{\rm thr}$.}
    \label{sc}
\end{figure}

To evaluate Eqs. (\ref{maineqb}) and (\ref{maineqf}), one needs  to  compute the  decay rates $\tau_\alpha^{-1}(q)$
for the fermionic and bosonic sectors and $q \lesssim T/u$.
Let us discuss the bosonic excitations first.
The simplest process of the bosonic decay obeying energy and momentum conservation is shown in the left panel of Fig.\ref{sc}. It corresponds to the decay of one boson mode into three and involves three bosons of the same chirality  (e.g. right) and one boson of the opposite chirality \cite{mp2013,bfduality}. 
The resulting decay rate of subthermal bosons  is given by   (see  Appendix \ref{ktblin}) we omit numerical factors)
\beqa
 \frac{1}{\tau_{B}(q)} \sim \begin{cases}
              \frac{\gamma q^{\frac{5}{3}}T^2}{u^5 l^{\frac{4}{3}}m_{*}^4}, \hspace{1cm} q<\frac{l^2 T^3}{u^3},\\[0.3cm]
            \frac{\gamma q^{2}T^3}{m_{*}^4u^6}, \hspace{1.2cm} \frac{l^2T^3}{u^3}<q<\frac{T}{u}.
      \end{cases}
      \label{nmt}
\eeqa
 Here $\gamma=\alpha^2 (1+\alpha)^2$  is a dimensionless interaction parameter related to the  LL parameter $K_0$ 
 by the relation $\alpha=\frac{1-K_0^2}{3+K_0^2}$.    The second line of Eq.~(\ref{nmt}) agrees with Ref.~\cite{mp2013}; the $q^{5/3}$ scaling as in the first line was obtained in the context of classical anharmonic chains \cite{prvz,lukspohn}.
 Note that the process shown in the left panel of Fig. \ref{sc} can be interpreted as either a contribution to the relaxation of one of the ``majority'' bosons 
 ($q$) or of the ``minority''  boson ($p$).    Kinematic constraints imply   $p\ll q$.  As a result, 
 the second contribution is exponentially suppressed at large momenta of the relaxing boson. It however dominates at small momenta and gives rise  to the first line in Eq.~(\ref{nmt}).

The process on the left panel of Fig. \ref{sc}  can be viewed as a decay of a right boson $q$ into two other right bosons assisted by a left mover $p$. The participation of the later is required by kinematic constraints on the Fermi golden-rule level. However, once the bosonic spectrum is broadened by some relaxation processes,  direct decay of a bosonic excitation into two bosons of the same chirality becomes possible. In particular, at $q<q_{\rm thr}$, where $q_{\rm thr}$ is a threshold momentum, 
\begin{equation}
q_{\rm thr}=\frac{T^{1/3}}{u m_{*}^{2/3} l^{4/3}}\equiv
\frac Tu\left(\frac{T_{\rm FB}}{T}\right)^{2/3},
\end{equation}
the {\it self-consistent} treatment of the process  shown in the right panel of Fig.~\ref{sc} provides the dominant contribution to the relaxation of bosons \cite{andreev80,sam,bovogangardt} (see Appendix \ref{ktblin}),
\beqa
 \tau_B^{-1}(q) \sim
 q^{3/2}m_{*}^{-1}\sqrt{T/u}, \hspace{2cm} q < q_{\rm thr}.
      \label{nmt00}
   \eeqa
 
 Summarising the above analysis, we get 
 \beqa
 \frac{1}{\tau_B(q)} \sim \begin{cases}
 q^{3/2}\frac{ \sqrt{T/u}}{m_{*}}, \hspace{.57cm} q < q_{\rm thr},\\[0.1cm]
  \frac{\gamma q^{\frac{5}{3}}T^2}{u^5 l^{\frac{4}{3}}m_{*}^4}, \hspace{1cm} q_{\rm thr}<q<\frac{l^2 T^3}{u^3},\\[0.3cm]
         \frac{\gamma q^{2}T^3}{m_{*}^4u^6},\hspace{1.2cm} \max\{\frac{l^2 T^3}{u^3}, q_{\rm thr}\} <q<\frac{T}{u}.\\
 \end{cases}
      \label{nmt1}
      \eeqa
The last two regimes in Eq. (\ref{nmt1}) can be absent if the corresponding momentum interval vanishes. Specifically, for $T<T_{\rm FB}$ the bosonic relaxation rate is always given by the first line in Eq.~(\ref{nmt1}). The intermediate regime,  $q_{\rm thr}<q<\frac{l^2 T^3}{u^3}$,  disappears at 
 $T<T_H=u^{\frac{3}{4}} l^{-\frac{5}{4}} m_{*}^{-\frac{1}{4}}$.

As for the fermionic excitations, their lifetime was discussed in Refs. \cite{imambekov11, bfduality,Khodas2007,Lunde2007}. It is given by
\beqa
\frac{1}{\tau_F(q)} \sim
                          \frac{\gamma l^4 T^7}{m_{*}^2 u^8},\hspace{1cm}q < \frac{T}{u}.
      \label{tautrfrm1}
\eeqa
Note that while the bosonic decay rate vanishes at $q\rightarrow 0$ limit, the fermionic rate remains finite.
This implies that the low-frequency behavior of the thermal conductivity is dominated by bosons.

We now calculate the real part of thermal conductivity as a 
 function of $\omega$ and T,  using  decay rates Eqs.(\ref{nmt1}) and (\ref{tautrfrm1}). 
 For $T<T_{\rm FB}$, we find 
\beqa
 \sigma^\prime(\omega) \sim \begin{cases}
 \frac{T^{\frac{11}{3}}l^4 u^{-\frac{5}{3}} m_{*}^{\frac{2}{3}}}{\omega^{\frac{1}{3}}}, \hspace{0.4cm}\omega<\frac{\gamma^3 T^{23}m_{*}^2l^{24}}{u^{20}},\\[0.2cm]
 \frac{u^5}{\gamma T^4l^4}, \hspace{1.5cm} \frac{\gamma ^3 T^{23}m_{*}^2l^{24}}{u^{20}}  < \omega <  \frac{\gamma l^4T^7}{m_{*}^2u^8},\\[0.2cm]
 \frac{1}{\omega^2} \frac{\gamma T^{10} l^4}{u^{11}m_{*}^4}, \hspace{1.0cm}
 \frac{\gamma l^4T^7}{m_{*}^2u^8}<\omega .    
\end{cases}
\label{conductivitylt}
\eeqa
For details of the calculations and results for $T >T_{\rm FB}$ see Appendix \ref{ftcht}.
To obtain the overall picture, the ``ballistic'' contribution (\ref{bal}) should be taken into account. At sufficiently high frequencies, $\omega\gg 1/\tau_{\rm U}$, the ballistic mode associated with the conservation of the momentum of bosonic excitations does not contribute to the real part of $\sigma(\omega)$ and  $\sigma(\omega)\approx \sigma^\prime(\omega)$.  At $\tau_{U}\omega \lesssim 1$ the   the ballistic channel of the energy propagation becomes gapped and contributes an exponentially large but frequency-independent constant $\sigma^{\rm bal} \simeq \pi u T\tau_U/3$ to the thermal conductance. 

The resulting behavior of $\sigma(\omega)$ is shown in Fig.\ref{tl} for $T <T_{\rm FB}$ and in Figs. 1 and 2 of  Appendix \ref{ftcht} for $T >T_{\rm FB}$. At $\omega<1/\tau_{\rm U}$, we observe a universal $\omega^{-1/3}$ scaling of $\sigma(\omega)$. 
This behavior can be traced back to the contribution of bosons with momentum $q\ll T/u$ that have the relaxation rate specified in first line of Eq. (\ref{nmt1}).  
It is consistent with the predictions of the fluctuating hydrodynamics.   This is to be expected as strongly subthermal bosonic modes  correspond to  classical density waves of the hydrodynamic theory.

\begin{figure}
   \begin{center}
  \includegraphics[width=8.75cm]{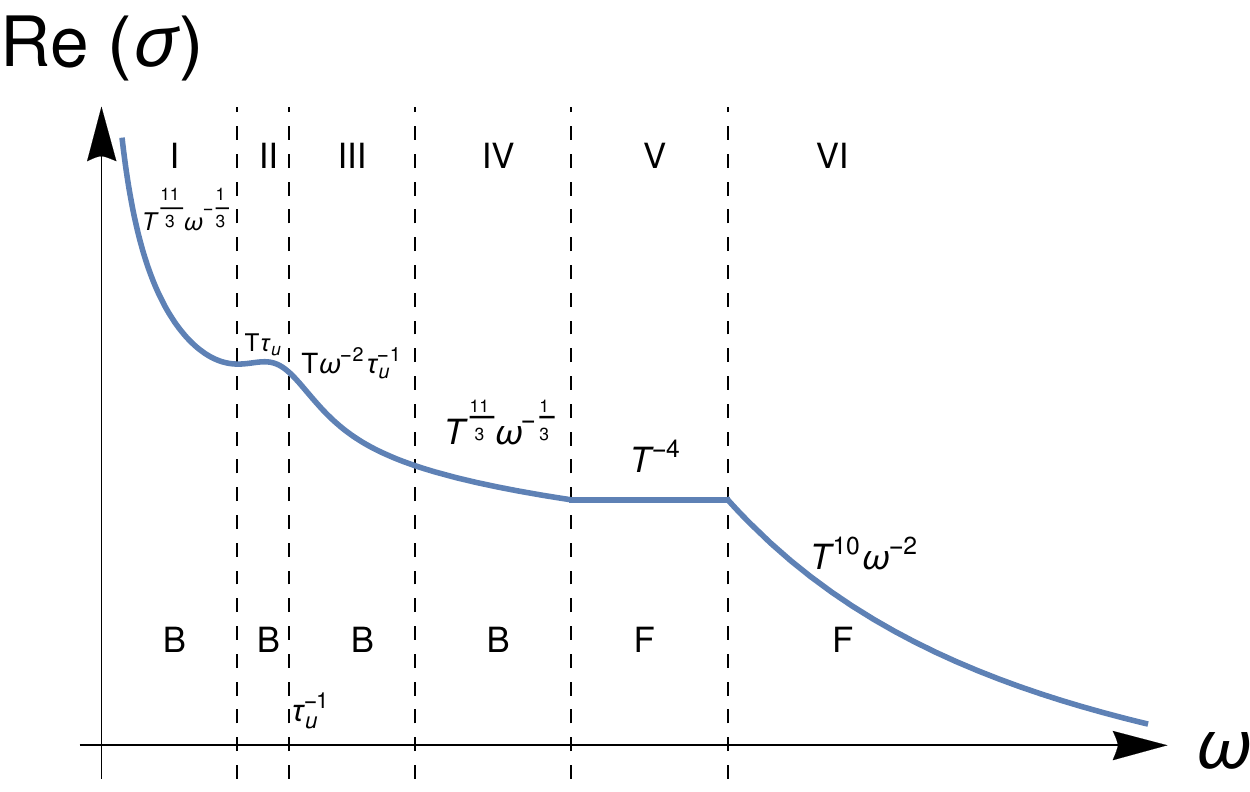}
   \end{center}
    \caption{${\rm Re}\:\sigma(\omega)$ at $T<T_{\rm FB}$. For each regime, the $T$ and $\omega$ scaling is shown.
    Label $F$ and $B$  indicates  whether the  dominant contribution comes from the fermionic or bosonic sector.  
    In region II and III  the contribution of the momentum zero mode (\ref{bal}) is dominant, while other regions are dominated by finite energy modes, Eq.(\ref{conductivitylt}). For lowest $\omega$, region I,  the $\omega^{-1/3}$ dependence translates into the $L^{1/3}$  scaling of $\sigma(L)$, analogous to that obtained for classical fluids \cite{nrs,Spohn2014}.}
    \label{tl}
\end{figure}

We now discuss the scaling of the DC  conductivity with the system size $L$.
In contrast to the frequency scaling, the contribution of the zero mode associated to the conservation of the bosonic momentum is always real. 
In fact at scales shorter than $L_{\rm U}$ the entire DC thermal conductance is dominated by this {\it ballistic} contribution,  $\sigma(L) =\pi T L/6$, with the other modes providing only subleading corrections.  The situation changes in the limit $L\gg L_{\rm U}$ where the contribution of the zero mode  becomes a size-independent constant, and non-zero modes start to be dominant. 

As was found above, the thermal conductivity $\sigma(\omega\rightarrow 0)$ is governed by bosons. 
In this limit, the bosonic lifetime diverges as  power law  $\tau^{-1}(q) \propto q^z$.
In our case $z=3/2$, that is consistent with KPZ scaling \cite{bovogangardt}.
The divergence of the life time implies  that bosons 
 with momentum below  $q<L^{-1/z}$ propagate through the system almost ballistically.
The contribution of these  bosons to the  thermal conductance can be estimated as 
 \beqa
 G(L)\sim  L^{-1/z}\,.
  \label{fr1}
 \eeqa
 This implies that for $z=3/2$ the conductivity scales as $\sigma(L)=L^{1/3}$.
Note that this is a manifestation of the  L{\'e}vy-flight character of the energy transport in the system, cf. Ref. \cite{Spohn2014}.
 To find the corresponding L{\'e}vy-flight distribution function,  one needs to relate $z$  to the L{\'e}vy-flight parameter $\alpha$. This can be done by comparing the diffusion coefficient $D_E$ of the L{\'e}vy-flight process with 
 the thermal conductivity computed above. The thermal conductance G  is related to the energy diffusion coefficient  as
$G= \sigma/L \sim D_E/L$.
 To estimate $D_E$  we compute 
   \beqa
& D_E=\frac{\langle x^2\rangle}{t}  \sim \int_0^{L/u} dx~ x^2 ~  x^{-\alpha -1}\sim L^{2-\alpha},
\label{DLevy}
\eeqa
where we used that the tails of the L{\'e}vy distribution function at time $t$ decay with distance as $x^{-\alpha -1} t$. 
Comparing Eqs. (\ref{fr1}) and (\ref{DLevy}), we obtain the following relation between the exponent $z$ controlling decay rate and $\alpha$ of L{\'e}vy flight,
$1-\alpha= -z^{-1}$, so that 
the  conductivity scales as
$\sigma \sim L^{2-\alpha} = L^{1-z^{-1}}$.
For $z= \frac{3}{2}$, the L{\'e}vy parameter $\alpha=5/3$ and  $\sigma(L)$  scales as $L^{\frac{1}{3}}$. 
Note that  the scaling in $L$  for  L{\'e}vy-flight regime  can be obtained from the scaling  
in  $\omega$ by   the replacement  $\omega \rightarrow u/L$.
The $L^{1/3}$ scaling  agrees with the one found  in classical fluids \cite{nrs,Spohn2014}. 

To summarize, we computed the thermal conductivity of 1D electronic fluid as a function of frequency $\omega$, temperature $T$, and system length $L$.
For energy scales  below  bosonic Umklapp time, the momentum of bosonic and fermionic fluids are separately conserved. 
The momentum zero mode give rise to the ballistic heat conductivity Eq.(\ref{bal}).
This corresponds to purely imaginary $\sigma(\omega)$ and results  in the LL thermal conductance
$\pi^2T/3h$\cite{kane96} for a finite sample. 
The massive modes of the fermionic and bosonic collision integrals contribute to the real part of the heat conductivity,
yielding subleading  in $1/L$ corrections to the thermal conductance.
However, they may be detected  via measuring real part  of $\sigma(\omega)$ at $\omega> u/L$.
The real part of $\sigma(\omega)$ exhibits several regimes.
For temperatures $T<T_{\rm FB}$ it is determined by fermionic modes for not too low frequencies, see regions V and VI in Fig.~\ref{tl}. 
 At the lowest frequencies, the conductivity is determined by low-momentum bosonic modes, yielding $\sigma(\omega) \propto \omega^{-1/3}$.
 The length dependence of the thermal conductance depends on the relation between $L$ and the bosonic umklapp length $L_U$.  For $L \ll L_U$,   the transport is ballistic, $\sigma(L) \propto L$, as expected for LL. On the other hand, for  $L \gg L_U$, we find $\sigma(L) \propto L^{1/3}$, as expected for  a classical fluid \cite{nrs,Spohn2014}.

We close by briefly discussing  prospective research directions.
First, while our computations were done within kinetic theory, these results can be found also within the hydrodynamic approach.
While for $L>L_U$ the hydrodynamic theory has three modes (particle-number, momentum, and energy conservation), for $L<L_U$ the number of modes is four, due to the additional zero mode discussed above. In both regimes, kinetic coefficients are expected to be anomalous. 
The hydrodynamic framework  is particularly  convenient for computing scaling functions describing heat conductivity and pulse evolution \cite{future}.
Second, whereas our computations were done for a strictly 1D  system, we expect that an  anomalous scaling of heat conductance should hold also for other low-dimensional quantum electronic fluids (quasi-1D and 2D geometries).

 {\it Note added:}  Shortly after our paper was submitted,  a related  preprint \cite{Matveev2019} appeared,
which addresses the same problem  solely within the  fermionic  approach.  
Results  of Ref.~\cite{Matveev2019} for $\sigma(\omega)$ match  ours in what concerns the  ``plateaus'' II and V   in  Fig. \ref{tl}
but do not capture other regions, where $\sigma(\omega)$ is controlled by a slow relaxation of bosonic modes.

\textit{ Acknowledgements:} 
 D. G. was supported
by ISF (grant 584/14) and Israeli Ministry of
Science, Technology and Space.

\begin{widetext}
\appendix
\section{Kinetic theory for bosonic excitations}
\label{ktb}
\subsection{Brief details of bosonization}
\label{bsnz}
In this section we briefly recap the bosonization approach for electrons with a finite curvature\cite{stone,delft98,gogolin,giamarchi,bfduality}. 
The system of interacting electrons in  one dimension is described by the microscopic Hamiltonian
\begin{equation}
 H=\sum_{\eta=R/L}\int dx \psi^+_\eta(x)\left(-i\eta v_F\partial_x-\frac{1}{2m}\partial_x^2\right)\psi_\eta(x)\\
 +\frac{1}{2}\int dx dx' g(x-x')\rho(x)\rho(x').
 \label{hm1}
\end{equation}
After the bosonisation procedure,  the Hamiltonian can be represented in  
 terms of the density field, 
\beqa
 H=\sum_{\eta}\int dx :\left(\pi v_F\rho_\eta^2(x)+\frac{2\pi^2} {3m}\rho_\eta^3(x)\right):_B
 +\frac{1}{2}\int dx dx' g(x-x'):\rho(x)\rho(x'):_B\,.
 \label{sec2:HamiltonianBoson0}
\eeqa
Here $::_B$ stands for normal ordering with respect to the bosonic modes.
The coupling between left and right chiral sectors can be eliminated up to a cubic level 
by performing unitary transformations, $R =U \rho_R U^+ ,\, L =U \rho_L U^+\,$,
see Ref.\cite{bfduality} for the details.  

After the rotation, the bosonic Hamiltonian reads
\begin{eqnarray}
 H &=&: \frac{\pi}{L}\sum_{q}u_q
 R_qR_{-q}
+\frac{1}{L^2}
\sum_{{\bf q}}  \Gamma^{B, RRR}_{\bf q} R_{q_1}R_{q_2}R_{q_3}
 +\frac{1}{L^3}\sum_{\bf q} \Gamma^{B, RRRL}_{\bf q}R_{q_1}R_{q_2}R_{q_3}L_{q_4}:_B+
\left(R\leftrightarrow L\right).
\label{sec2:HamiltonianBosonFinal}
\end{eqnarray}
The  bosonic vertices  $\Gamma_{\bf q}$  in  Eq.(\ref{sec2:HamiltonianBosonFinal}) in the low momentum 
limit  ($ql \ll 1$)  are  
\begin{eqnarray}
 \Gamma^{B, RRR}_{\bf q}&=&\frac{2\pi^2}{3 m_* }\left(1-\frac{\alpha l^2}{2}
(q_1^2+q_2^2+q_3^2)\right)\,,\,\,
 \Gamma^{B, RRRL}_{\bf q}=\frac{4 \pi ^3 \alpha}{3 m_{*}^2 u} 
\left[1-\frac{3 \alpha }{2}+\frac{15}{4}l^2\frac{q_1 q_2 q_3}{q_4}\right].
 \label{sec2:GammaB_RRR}
\label{sec2:GammaB_RRL}
\end{eqnarray}
Here  $m_{*}=\frac{4\sqrt{K_0}}{3+K_0^2} m$ is renormalised  mass of the electron,  $\alpha=\frac{1-K_0^2}{3+K_0^2}$  is dimensionless interaction strength, and $K_0$ is LL parameter.
Next, we employ the Hamiltonian (\ref{sec2:HamiltonianBoson0}) to derive a kinetic equation for the bosonic distribution function.

\subsection{Linearization of kinetic equation for bosons}
\label{ktblin}
The Fourier components of the densities $R(x)$ and $L(x)$ can be identified with
bosonic creation and annihilation operators via 
\begin{eqnarray}
R_q=\sqrt{\frac{L |q|}{2\pi}}\left(\Theta(q)b_q+\Theta(-q)b^+_{-q}\right),\nonumber\\
L_q=\sqrt{\frac{L |q|}{2\pi}}\left(\Theta(-q)b_q+\Theta(q)b^+_{-q}\right).
\label{sec3:BosonicOperators}
\end{eqnarray}

The bosonic distribution $N_B(q,x,t)$ is defined as
\beqa
N_B(q,x,t)= \frac{1}{2\pi}\int_{-\infty}^{\infty} d(q_1 -q_2) e^{i (q_1-q_2) x}  \big{\langle} b_{q_1}^+(t) b_{q_{2}}(t)\big{\rangle},
\eeqa
where $q=(q_1+q_2)/2$ and the operators $b_q$ and $b_q^+$ are defined in Eq.(\ref{sec3:BosonicOperators}).
Using a standart Keldysh formalism\cite{rammer,kamenev} one derives the kinetic equation for bosons
\beq
\frac{\partial N_{q}}{\partial t} +u_q \frac{\partial N_{q}}{\partial x} 
=\mathcal{I}_\text{out}\bigl[N_{q}\bigr] 
+ \mathcal{I}_\text{in}\bigl[N_{q}\bigr].
\label{e6}
\eeq
Here $u_q=\frac{d \omega_q}{dq}$ is the sound  velocity;  the incoming and outgoing parts of  the collision integral are  
\Beqa
\mathcal{I}_\text{out}\bigl[N_{q}\bigr] 
&=&  -\sum_{p,q_1,q_2}\!W_{q,p;q_1, q_2}
N_{q}N_{p}(1+N_{q_1})(1+N_{q_2}),
\\
\mathcal{I}_\text{in}\bigl[N_{q}\bigr] 
&=& 
\hspace{.4cm}\sum_{p,q_1,q_2}\!
W_{q,p;q_1,q_2}
(1+N_{q})(1+N_{p})N_{q_1}N_{q_2}.
\Eeqa
The matrix element
\beqa
W_{q_1^\prime, q_2^\prime;q_1^\prime, q_2^\prime}
&=&\frac{\gamma}{m_{*}^4 u^2} \ |q_1 q_2 q_1^\prime q_2^\prime|
\delta_{q_1^\prime + q_2^\prime,q_1^\prime + q_2^\prime}
\delta(\omega_{q_1}+\omega_{q_2}-\omega_{q_1^\prime}-\omega_{q_2^\prime}),
\label{e7}
\eeqa
where we defined 
$\gamma= \alpha^2 (1+\alpha)^2$.\\
If the deviations from the equilibrium  are  small, one can linearise the collision integral. 
It is conveniently done in the following parametrization:
\beq
N_q = n_q+g_q f_q,
\quad
g_q = \sqrt{n_q(1+n_q)\,}. 
\label{gdef}
\eeq
Here $f_q$ represents the deviation of the distribution function from the local equilibrium distribution $n_q$.
The linearised kinetic equation reads
\beqa
\frac{\partial f_{q}}{\partial t}+ \frac{u_q}{g_q} \frac{\partial n_{q}}{\partial T} \nabla T
=\!-\frac{\gamma}{m_{*}^4 u^2}\sum_{p,q_1,q_2}|q p q_1 q_2|\!
\left(\frac{f_q}{g_q} + \frac{f_p}{g_p} - \frac{f_{q_1}}{g_{q_1}} - \frac{f_{q_2}}{g_{q_2}}\right)
g_p g_{q_1} g_{q_2} 
\delta(\omega_{q}+\omega_{p}-\omega_{q_1}-\omega_{q_2})\,\delta_{q+p,q_1+q_2}.
\label{e10} 
\\
\nonumber
\eeqa
It is convenient to define a dimensionless momentum  $x=\frac{u}{2 \pi}\frac{q}{T}$.
We now expand Eq.(\ref{e10}) keeping the lowest order in $lT/u$.
After  Fourier transforming with respect to time and implementing  the delta functions in Eq.(\ref{e10}) one finds
\beqa
i \omega f(x) + S(x)=\int_{-\infty}^\infty\!dy\,\mathcal K(x,y) f(y).
\label{eqm1}
\eeqa
Here the source term
\beqa
S(x)= \frac{u}{T} \nabla T \frac{\pi x}{\sinh(\pi |x|)}\left(1  +\frac{4\pi^2 l^2 T^2 }{u^2} x^2[\pi x \coth \pi x-4] \right)\label{source}\,.
\eeqa 
Note that the source is an odd function,  $S(-x)=-S(x)$,  and therefore has a finite component along the momentum zero mode and no component along the energy zero mode. For this reason the momentum zero mode is essential for  low energy transport.
The  kernel $\mathcal{K}(x,y)$ is composed of  local  and  non local parts
 \beqa
 \mathcal{K}(x,y)= \mathcal{K}_{1}(x)\delta(x-y) + \mathcal{K}_{2}(x,y).
 \label{kerneld}
 \eeqa
The local part is  
\beqa
\mathcal K_{1}(x) &=
\begin{cases}
 \frac{\gamma T^{11/3} x^{5/3} l^{-4/3}}{m_{*}^4 u^{20/3}}\,, \hspace{0.7cm} x \ll \frac{l^2 T^2}{u^2},\\
 \frac{\gamma T^5}{m_{*}^4 u^8} \frac{1}{6}x^2(x^2 + 1)\,, ~~x > \frac{l^2 T^2}{u^2}.
\label{local0}
\end{cases}
\eeqa  
The non local part  of the collision integral  
\beqa
\mathcal K_{2}(x,y) &= \theta(xy) \frac{\gamma T^5}{m_{*}^4 u^8}\left(\frac{xy(x+y)}{\sinh\bigl[\pi(x+y)\bigr]}
- \frac{xy(x-y)}{\sinh\bigl[\pi(x-y)\bigr]}\right)\, + \theta(-xy) \mathcal H(x,y).\nn
\label{nonlocal}
\eeqa
Here 
\beqa
& \mathcal H(x,y) =\frac{\gamma T}{l^4 m_{*}^4 u^4} 
\frac{x^2 g_{q_1^\prime} g_{q_2^\prime}}{|y|\sqrt{y^2
+\frac{2 u^2 x}{3 l^2 \pi^2 T^2 |y|}}}+&\frac{\gamma T}{l^4 m_{*}^4 u^4} 
\frac{x^2 g_{q_1} g_{q_2}}{|y|\sqrt{y^2
-\frac{2 u^2 x}{3 l^2 \pi^2 T^2 |y|}}}~\theta \left(|y|- \left(\frac{2 u^2  x}{3l^2 \pi^2 T^2}\right)^{\frac{1}{3}}\right)+ 
(x \leftrightarrow y),\nn
\eeqa
where the function $g_x\equiv\frac{1}{2 \sinh(\pi |x|)}$
and
$q_{1,2}=\frac{1}{2} \left(y \pm \sqrt{y^2
-\frac{2 u^2 x}{3 l^2 \pi^2 T^2 |y|}}\right), \,
q_{1,2}^\prime=\frac{1}{2} \left(\pm y- \sqrt{y^2
+\frac{2 u^2 x}{3 l^2 \pi^2 T^2 |y|}}\right)$.
\newline
Employing  Eqs.(\ref{kerneld},\ref{local0}), on the level of diagonal approximation, we find the bosonic decay rates 

\beqa
 \frac{1}{\tau_{B}(q)}=\begin{cases}
              \frac{\gamma q^{\frac{5}{3}}T^2}{u^5 l^{\frac{4}{3}}m_{*}^4},\hspace{1.1cm}q<\frac{l^2 T^3}{u^3},\\\\
            \frac{\gamma q^{2}T^3}{m_{*}^4u^6}, \hspace{1.3cm}\frac{l^2T^3}{u^3}<q<\frac{T}{u},\\\\
             \frac{\gamma q^{4}T}{m_{*}^4u^4},\hspace{1.5cm}q>\frac{T}{u}.\\
      \end{cases}
      \label{anmt}
\eeqa
The bosonic decay  rates (\ref{nmt}) have been computed in Refs.\cite{prvz,mp2013,bfduality}.
The analytic computation of a decay rate with the full kernel Eq.(\ref{kerneld})  is very challenging. 
However, the numerical analysis we performed showed that results for the decay rate in the small momentum limit found 
within the full collision integral and those computed  within the diagonal approximation agree. 

Now we discuss the bosonic  vertices $\Gamma^{B, RRR}_{\bf q}$ and $\Gamma^{B, LLL}_{\bf q}$,
that  describe  three boson interaction Eq.(\ref{sec2:GammaB_RRR}).
The treatment of these vertices on the golden rule level is subtle, and for the bosonic spectrum in the absence of broadening the answer is ill-defined.   However, if  the broadening of the spectrum is computed self-consistently\cite{andreev80,sam} 
one finds  

\beqa
\frac{1}{\tau_{s}(q)} =\begin{cases}
            q^{3/2}\frac{ \sqrt{T/u}}{m}, \hspace{1cm}q <q_{\rm thr},\\
              0, \hspace{2.35cm}q>q_{\rm thr}.\\
      \end{cases}
      \label{qcond0}
\eeqa
Here the value of the threshold  momentum $q_{\rm thr}= \frac{T^{1/3}}{u m^{2/3} l^{4/3}}$  is determined by the condition that the energy level broadening (\ref{qcond0}) exceeds the nonlinear correction $u l^2 q^3$ to the bosonic dispersion relation at momentum $q$.
This reproduces the result  found in Ref.\cite{andreev80,sam,bovogangardt}.
\subsection{Computation of the thermal conductivity}
\label{btc}
Now we turn to computation of the  thermal conductivity.
We decompose the source S(x) term in Eq.(\ref{eqm1}) into a part parallel  
\beqa
S_{||}= \frac{\pi  |x|}{\sinh(\pi x)}  \frac{u\nabla T  }{ T },\nn
\eeqa
and perpendicular   to  the momentum zero mode
\beqa
S_{\perp}=\frac{16 \pi^3l^2T\nabla{T}}{5u} \frac{ |x|(1-5x^2)}{ \sinh(\pi x)}\,.
\eeqa
Similarly we decompose the sought after solution $f$ into $f_{||}$ and $f_{\perp}$.  Within the relaxation-time approximation the  solution of the bosonic kinetic equation  is  given by
 \beqa
 f_{||}= - \frac{S_{||}}{i \omega}\,\,\,
{\rm and} \,\,\,  
f_{\perp} =\frac{S_{\perp}}{\tau_B^{-1}(q)-i \omega }.
\label{fperp}
\eeqa
We have checked 
 that this approximate result  is in excellent  agreement with the exact solution obtained by inverting the full collision integral operator numerically.\\ 
 The ``longitudinal'' part of the correction to the distribution function, $f_{||}$, gives rise to the ballistic contribution to the thermal conductance, 
 Eq.(\ref{bal})of the main text.
Plugging $f_{\perp}$ from Eq.(\ref{fperp}) into the formula for thermal current,  we find the real part of thermal conductivity carried by  the bosonic excitations
\beqa
&{\rm Re} \, \sigma^B(\omega)\simeq \frac{T^4 l^4}{u^2} {\rm Re}\left[\int_0^{T/u} ~   \frac{(d q)}{{\tau^{-1}_{B}(q)}-i \omega }\right]\,.
\label{amaineqb}
\eeqa
The   momentum  integration in Eq.(\ref{amaineqb})  is cut by  the  value of the thermal momentum $T/u$, because 
beyond it  the integrand  is exponentially suppressed.
By comparing the self-consistent one-into-two   boson decay rate with  one-into-three  boson decay rate 
one observes that the first one is faster, for $q<q_{\rm thr}$.
Therefore, the integration over momentum in Eq.(\ref{amaineqb}) is divided into the region $q \gtrless q_{\rm thr}$.
For $T< T_{\rm FB}$  it implies  that  $q_{\rm thr} > T/u$ and the thermal conductivity  alway dominated by 
one into two bosons decay processes, resulting in  
\beqa
 {\rm Re}  \, \sigma^B(\omega)\simeq
              \frac{ T^{\frac{11}{3}} l^{4}  u^{-\frac{5}{3}} m_{*}^{\frac{2}{3}} }{\omega^{\frac{1}{3}}}, \hspace{1cm}\omega <\frac{T^2}{m_*u^2}.
 \label{cs3c}
\eeqa
 For $  T_{\rm FB}<T<T_H$ where $T_H=u^{\frac{3}{4}} l^{-\frac{5}{4}} m_{*}^{-\frac{1}{4}}$ the thermal conductivity  Eq.(\ref{amaineqb}) is given by a sum
 of parts coming from below  and above $q_{\rm thr}$. ${\rm Re}(\sigma^B)=I_1(\omega)+I_2(\omega)$.
The contribution coming from momenta below $q_{\rm thr}$ is given by 
\beqa
I_1(\omega) =  \frac{ T^{\frac{11}{3}} l^{4}  u^{-\frac{5}{3}} m_{*}^{\frac{2}{3}} }{\omega^{\frac{1}{3}}} .
\eeqa
The contribution coming from momenta above $q_{\rm thr}$ is given by 
\beqa
I_2(\omega) = \begin{cases}
\frac{\gamma l^4 T^{10}}{m_{*}^4 u^{11}} \frac{1}{\omega^2}, \hspace{2.2cm} \omega >\frac{\gamma T^5}{m_{*}^4 u^8},\\\\
 \frac{ T^{\frac{5}{2}} m_{*}^{2}  u l^4  }{\omega^{\frac{1}{2}}\gamma^{\frac{1}{2}}}, \hspace{2.2cm} \frac{\gamma T^{\frac{11}{3}}}{u^8 m_{*}^{\frac{16}{3}} l^{\frac{8}{3}}} < \omega <\frac{\gamma T^5}{m_{*}^4 u^8}, \\\\
  T^{\frac{2}{3}} m_{*}^{\frac{14}{3}} l^{\frac{16}{3}} u^5 \gamma^{-1},\hspace{1.05cm}\omega<\frac{\gamma T^{\frac{11}{3}}}{u^8 m_{*}^{\frac{16}{3}} l^{\frac{8}{3}}}.\\
 \end{cases}
\eeqa
Thus, the bosonic contribution to the thermal conductivity   is given by
\beqa
 {\rm Re}\, \sigma^B(\omega)=\begin{cases}
              \frac{ T^{\frac{11}{3}} l^{4}  u^{-\frac{5}{3}} m_{*}^{\frac{2}{3}} }{\omega^{\frac{1}{3}}}, \hspace{1.65cm} \omega <
              \frac{\gamma^3 T^9}{m_*^{12}u^{20}l^4},
           \\\\
              T^{\frac{2}{3}} m_{*}^{\frac{14}{3}} l^{\frac{16}{3}} u^5 \gamma^{-1}, \hspace{1cm} 
                \frac{\gamma^3 T^9}{m_*^{12}u^{20}l^4}
              <\omega<\frac{\gamma T^{\frac{11}{3}}}{u^8 m_{*}^{\frac{16}{3}} l^{\frac{8}{3}}},\\\\
               \frac{ T^{\frac{5}{2}} m_{*}^{2}  u l^4  }{\omega^{\frac{1}{2}}\gamma^{\frac{1}{2}}}, \hspace{2.2cm}  \frac{\gamma T^{\frac{11}{3}}}{u^8 m_{*}^{\frac{16}{3}} l^{\frac{8}{3}}} < \omega <\frac{\gamma T^5}{m_{*}^4 u^8},\\\\
             \frac{\gamma l^4 T^{10}}{m_{*}^4 u^{11}} \frac{1}{\omega^2}, \hspace{2.3cm}  \omega >\frac{\gamma T^5}{m_{*}^4 u^8}.\\
             \end{cases}
 \label{cs1c}
\eeqa
For $T \gg T_H$, the thermal conductivity  Eq.(\ref{amaineqb}) is given by a sum
 of three parts : $q< q_{\rm thr}$,  $q_{\rm thr}<q<\frac{l^2 T^3}{u^3}$ and  $q>\frac{l^2 T^3}{u^3}$ . Thus,
  ${\rm Re}\, \sigma^B(\omega)=I_1(\omega)+I_2(\omega) +I_3(\omega)$.
Adding these three contributions, we thus have
\beqa
 {\rm Re}\, \sigma^B(\omega)=\begin{cases}
              \frac{ T^{\frac{11}{3}} l^{4}  u^{-\frac{5}{3}} m_{*}^{\frac{2}{3}} }{\omega^{\frac{1}{3}}}, \hspace{1.8cm}\omega <\frac{\gamma^3 T^{\frac{17}{3}} l^{-\frac{20}{3}} u^{-16}}{m_{*}^{\frac{34}{3}}},\\\\
                          T^{\frac{16}{9}} m_{*}^{\frac{40}{9}} l^{\frac{56}{9}} u^{\frac{11}{3}} \gamma^{-1}, \hspace{.9cm}{\scriptstyle \frac{\gamma^3 T^{\frac{17}{3}} l^{-\frac{20}{3}} u^{-16}}{m_{*}^{\frac{34}{3}}}<\omega<\frac{\gamma T^{\frac{23}{9}}}{u^{\frac{20}{3}} m_{*}^{\frac{46}{9}} l^{\frac{32}{9}}}},\\\\
          \frac{ T^\frac{14}{5} l^\frac{24}{5} m_{*}^\frac{12}{5} u}{ \omega^{2/5}\gamma^{3/5}               }, \hspace{1.8cm} \frac{\gamma T^{\frac{23}{9}}}{u^{\frac{20}{3}} m_{*}^{\frac{46}{9}} l^{\frac{32}{9}}}<\omega<\frac{\gamma l^7 T^{12}}{m_{*}^4 u^{15}},\\\\
           \frac{ l^2 u^7 m_{*}^4}{ T^2 \gamma             }, \hspace{2.7cm}\frac{\gamma l^7 T^{12}}{m_{*}^4 u^{15}}<\omega<\frac{\gamma l^4 T^9}{m_{*}^4 u^{12}},\\\\
               \frac{ T^{\frac{5}{2}} m_{*}^{2}  u l^4  }{\omega^{\frac{1}{2}}\gamma^{\frac{1}{2}}}, \hspace{2.4cm}\frac{\gamma l^4 T^9}{m_{*}^4 u^{12}} < \omega <\frac{\gamma T^5}{m_{*}^4 u^8},\\\\
             \frac{\gamma l^4 T^{10}}{m_{*}^4 u^{11}} \frac{1}{\omega^2}, \hspace{2.4cm}\frac{\gamma T^5}{m_{*}^4 u^8}<\omega. \\
             \end{cases}
 \label{csht}
\eeqa
 
\section{Kinetic theory for fermionic excitations}
\label{tau1f}
\subsection{Refermionization}
\label{rfrm}
The bosonized Hamiltonian (\ref{sec2:HamiltonianBosonFinal}) can be recast in terms of fermionic operators as 
\begin{equation}
 R_q=\sum_{k}c^\dagger_{R, k}c_{R, k+q}\,, \qquad L_q=\sum_{k}c^\dagger_{L, k}c_{L, k+q}.
\label{sec4:Fermions}
 \end{equation}
This results  in a refermionised  Hamiltonian
\begin{eqnarray}
 H &=& \sum_{k}\epsilon_{R, k} :c^\dagger_{R, k}c_{R, k}:_F + \frac1L\sum_{\bf k}\Gamma^{F, RR}_{\bf k}:c^\dagger_{R, k_1}c^\dagger_{R, k_2}c_{R,
k_2^\prime}c_{R, k_1^\prime}:_F \nonumber\\
  &+& \frac1L\sum_{\bf k}\Gamma^{F, RL}_{\bf k}:c^\dagger_{R, k_1}c^\dagger_{L, k_2}c_{L,
k_2^\prime}c_{R, k_1^\prime}:_F + \frac{1}{L^2}\sum_{\bf k}\Gamma^{F, RRR}_{\bf k}:c^\dagger_{R, k_1}c^\dagger_{R,
k_2}c^\dagger_{R, k_3}c_{R, k_3^\prime}c_{R, k_2^\prime}c_{R, k_1^\prime}:_F \nonumber
\\
  &+& \frac{1}{L^2}\sum_{\bf k}\Gamma^{F, RRL}_{\bf k}:c^\dagger_{R, k_1}c^\dagger_{R,
k_2}c^\dagger_{L, k_3}c_{L, k_3^\prime}c_{R, k_2^\prime}c_{R, k_1^\prime}:_F +  (R\longleftrightarrow L). 
\label{sec4:HamiltonianFermionsFinal}
\end{eqnarray}
 We denote by  ${\bf k}$ in each of vertices $\Gamma^{F, \ldots}_{\bf
k}$ the full set of all momenta of the fermionic operators involved. As shown in Ref.\cite{bfduality} the dominant vertex for energy relaxation is $\Gamma^{F, RRL}_{\bf k}$.
 In the bosonic description it  corresponds to 1 boson going into 3 bosons scattering process, e.g. Eq.(\ref{sec2:GammaB_RRL}).
\begin{eqnarray}
 \Gamma^{F, RRL}_{\bf k} &=& \frac{5 \alpha 
l^2\pi^2(k_1-k_2)(k_1^\prime-k_2^\prime)}{16m^{*2}u(k_3-k_3^\prime)} \times \left[(k_1-k_2)^2-(k_1^\prime-k_2^\prime)^2\right].
 \label{sec4:GammaFRRL}
 \end{eqnarray}

%

\subsection{Kinetic equation for composite fermions}
\label{ktf}
Using the Keldysh formalism, one can derive a kinetic equation for the fermionic distribution $N_k (x,t)$ 
\beqa
N(k,x,t)= \int_{-\infty}^{\infty} \frac{d(k_1 -k_2)}{2\pi} e^{i (k_1-k_2) x}  \big{\langle} c_{k_1}^+(t) c_{k_{2}}(t)\big{\rangle},
\eeqa
where $k = \frac{k_1+k_2}{2}$ and the operators $c_k$ and $c_k^+$ are defined in Eq.(\ref{sec4:HamiltonianFermionsFinal}).
By repeating the standard steps of Keldysh formalism with the Hamiltonian (\ref{sec4:HamiltonianFermionsFinal})
 one derives the kinetic equation for composite fermions
\beqa
\frac{\partial N_k}{\partial t} +v_{k} \frac{\partial N_k }{\partial x}= I [N_k].
\label{blff}
\eeqa
Here $v_k = \partial_k \epsilon_k$ is a velocity of fermions,
$ \epsilon_k= \frac{k^2}{2m_*}-\frac{k_F^2}{2m_*}$, and  $k_F= m_{*} u$.
The  fermionic collision integral $I [N]$ is given by
\beq
I [N]=I_\text{out}\bigl[N\bigr] 
+ I_\text{in}\bigl[N\bigr].
\eeq
Here 
\Beqa
I_\text{out}\bigl[N\bigr]_k 
&=&  -\sum_{k_2,k_3,k^\prime,k_2^\prime,k_3^\prime}W^{k^\prime k_2^\prime k_3^\prime}_{k k_2 k_3}
N_{k}N_{k_2}N_{k_3}(1-N_{k^\prime})(1-N_{k_2^\prime})(1-N_{k_3^\prime})
\Eeqa
is outgoing 
\Beqa
I_\text{in}\bigl[N\bigr] _k
&=& \sum_{k_2,k_3,k^\prime,k_2^\prime,k_3^\prime}W^{k^\prime k_2^\prime k_3^\prime}_{k k_2 k_3}
(1-N_{k})(1-N_{k_2})(1-N_{k_3})N_{k^\prime}N_{k_2^\prime}N_{k_3^\prime}
\Eeqa
and incoming parts.
The matrix element of three fermion collision\cite{bfduality} is given by
\beqa
&W^{k^\prime k_2^\prime k_3^\prime}_{k k_2 k_3}= \frac{\gamma l^4}{m_{*}^2 u} 
(k_2-k)^2 (k_2^\prime- k^\prime)^2~\delta( k_2 +k -k _2^\prime -k ^\prime)~ \delta(k_3 -k_3^\prime) \nn.
\eeqa
Near the equilibrium one can  linearize  the collision integral using the  ansatz
\beqa
N=n+g f\,.
\eeqa
Here $n$ denotes the local equilibrium Fermi-Dirac distribution $n_k=\frac{1}{e^{\epsilon_k/T}+1}$ and $g_k=\sqrt{n_k(1-n_k)}=\frac{1}{2 \cosh{\epsilon_k/2 T}}$. \\
After Fourier transforming it in time, the linearized Boltzmann equation reads 
\beqa
-i \omega f_k +B^F_k \frac{\nabla T}{T^2} =  {\cal{I}} [f]_k,
\label{bfm}
\eeqa
where 
\beqa
 B^F_k=v_k \epsilon_k g_k .
 \label{bdef}
 \eeqa
and the linearized collision integral $\cal{I}$ is 
\beqa
&{\cal{I}}[f]_k=\frac{\gamma T l^4}{m_{*}^2 u^2}\int_{-\infty}^\infty dk_2   \int_{-\infty}^\infty dk^\prime \int_{-\infty}^\infty  dk_2^\prime  ~ (k-k_2)^2 (k^\prime- k_2^\prime)^2 \delta(k +k_2 -k^\prime -k_2^\prime)\nn\\
& g(k_2) g(k^\prime) g(k_2^\prime)\left(\frac{f(k)}{g(k)}+\frac{f(k_2)}{g(k_2)}-\frac{f(k^\prime)}{g(k^\prime)}-\frac{f(k_2^\prime)}{g(k_2^\prime)}\right).
\eeqa
On the level of diagonal approximation, we find the decay rate  
\beqa
&\tau_F^{-1}(k)=\frac{\gamma T l^4}{g(k)m_{*}^2 u^2} \int_{-\infty}^\infty dk_2   \int_{-\infty}^\infty dk^\prime \int_{-\infty}^\infty  dk_2^\prime  ~ (k-k_2)^2 (k^\prime- k_2^\prime)^2 \delta(k +k_2 -k^\prime -k_2^\prime) g(k_2) g(k^\prime) g(k_2^\prime).
\eeqa
This yields fermionic life time
\beqa
 \frac{1}{\tau_F(q)} =\begin{cases}
             \frac{\gamma l^4 T k^6}{m_{*}^2 u^2}, \hspace{2cm}k > \frac{T}{u},\\ \\
              \frac{\gamma l^4 T^7}{m_{*}^2 u^8}, \hspace{2.2cm}k < \frac{T}{u}.\\
      \end{cases}
      \label{fdecay}
      \eeqa
reproducing the results of Ref. \cite{imambekov11, bfduality,Khodas2007,Lunde2007}. We have also numerically studied the life time employing full kernel and found that  the results (\ref{fdecay}) hold.

Next, we compute the thermal conductivity in the fermionic channel.
Since fermions are massive particles, one should impose a zero momentum transfer condition\cite{Landau_SP1}.
It is achieved by subjecting the system to the gradient  of chemical potential $\nabla \mu$.
 After this procedure, we are left with ${{\cal B}}^F$ that is a part of $B^F$ orthogonal to the {\it momentum} zero mode
$\phi_p=\frac{1}{\sqrt{2 m_* T k_F}}~ k g_k$.

In the orthogonal subspace,  we now further decompose it into parts parallel  and perpendicular to the {\it number} zero mode
$\phi_n=\sqrt{\frac{k_F}{2 m_* T}} ~{\rm sgn}(k) g_k$.
This results in  the decomposition 
 ${\cal{B}}^F={{\cal B}^F_{||}}+{\cal B_{\perp}^F}$.
To  leading order in T,
\beqa
&{{\cal B}^F_{||}}=\frac{\sqrt{2} k_F^{\frac{5}{2}} T^{\frac{1}{2}}}{m_*^{\frac{3}{2}}} (\phi_p -\phi_n),\hspace{2cm}{{\cal B}^F_{\perp}}=B^F -\frac{\sqrt{2 m_*}  \pi^2 T^{\frac{5}{2}}}{3 k_F^{\frac{3}{2}}} \phi_p -  {{\cal B}^F_{||}}.
\label{bperpf}
\eeqa
We now substitute this decomposition into Eq.(\ref{bfm}) and compute the thermal conductivity.
\subsection{Computation of the thermal conductivity}
\label{frtc}
Within the diagonal approximation one can solve  the Boltzmann equation Eq.(\ref{bfm})
\beqa
f_{\rm bal} = \frac{{\cal B^F_{||}}}{-i \omega} \frac{\nabla T}{T^2},~~ f_\perp =\frac{\nabla T}{T^2} \frac{{\cal B^F_{\perp}}}{\tau_F^{-1} -i \omega}\,.
\label{fsolf}
\eeqa

Thus, the ballistic part of the thermal current is given by
\beqa
&J_{\rm bal}=\frac{\pi}{3} \frac{  T u}{i \omega}\nabla T \nn.
\eeqa
The ballistic part of the thermal current matches the one found within the bosonic approach.
This happens because in both descriptions it is  fully controlled by a corresponding (bosonic and fermionic) 
momentum  zero modes.

The real part of thermal conductivity is thus given by 
\beqa
{\rm Re}\; \sigma^F (\omega)\simeq\frac{T^2}{m_{*}^2u^2} {\rm Re}\int_0^{\frac{T}{u}} ~   \frac{(d q)}{{\tau_{F}(q)}^{-1} -i \omega }.
\label{fers}
\eeqa

Employing Eq.(\ref{fdecay}) for the fermionic life time, one finds 
 \beqa
& \displaystyle {\rm Re}\; \sigma^F (\omega)  = \begin{cases}
 \frac{1}{\omega^2} \frac{\gamma T^{10} l^4}{u^{11}m_{*}^4} \ , & \qquad \omega >  \frac{\gamma l^4T^7}{m_{*}^2u^8}  \ \ \text{(regime F1)}  \\
 \frac{u^5}{\gamma T^4l^4} \ , & \qquad \omega <  \frac{\gamma l^4T^7}{m_{*}^2u^8}  \ \ \text{(regime F2)}. 
 \end{cases}
\label{fcon}
\eeqa
\section{Final thermal conductivity: bosons vs fermions}
\label{ftc}
As we showed above, there are two channels of the energy transport in 1D  electronic fluid: fermionic and bosonic. The resulting contribution to thermal conductivity is given by a sum of bosonic and fermionic parts. For the imaginary part of the thermal conductivity, both fermionic and bosonic channels yield identical results,
and therefore one may use either description. 
For the real part of the thermal conductivity, this is not the case.
However, since ${\rm Re}~ \sigma \sim \tau$,  it is automatically determined by the long-living species. Therefore, up to a numerical prefactor,  the result can be obtained by summing both contributions.
Now we analyze the results for different temperature limits. 
\subsection{High temperature regime, $T>T_{\rm FB}$}
\label{ftcht}
For  $T_{\rm FB}<T$,  the bosonic description is suitable, $\tau_B(q)> \tau_F(q)$,  for all accessible momenta, $q<T/u$. 
Therefore  the real part of thermal conductivity is determined by $\sigma_B(\omega)$.

For $T_{\rm BF} <T <T_H$, $\sigma_B(\omega)$ is given by Eq.(\ref{cs1c}), thus leading to the total thermal conductivity
\beqa
\label{conductivityht}
{\rm Re}~ \sigma'(\omega)= \begin{cases}
 \frac{T^{\frac{11}{3}}l^4 u^{-\frac{5}{3}} m_{*}^{\frac{2}{3}}}{\omega^{\frac{1}{3}}}, \hspace{1.5cm} \omega<\frac{\gamma^3 T^9}{m_*^{12}u^{20}l^4},  \\
    T^{\frac{2}{3}} m_{*}^{\frac{14}{3}} l^{\frac{16}{3}} u^5 \gamma^{-1}, \hspace{.8cm} {\scriptstyle
    \frac{\gamma^3 T^9}{m_*^{12}u^{20}l^4}
    <\omega<\frac{\gamma T^{\frac{11}{3}}}{u^8 m_{*}^{\frac{16}{3}} l^{\frac{8}{3}}}},\\

  \frac{T^{5/2}m_{*}^2ul^4}{\omega^{\frac{1}{2}} \gamma^{\frac{1}{2}}},\hspace{1.9cm} \frac{\gamma T^{\frac{11}{3}}}{u^8 m_{*}^{\frac{16}{3}} l^{\frac{8}{3}}} < \omega <\frac{\gamma T^5}{m_{*}^4 u^8},\\ \\

    \frac{\gamma T^{10}l^4}{\omega^2 u^{11}m_{*}^4}, \hspace{2.2cm} \omega >  \frac{ \gamma T^5}{m_{*}^4 u^8}.\\       
\end{cases}
\eeqa
These results lead to the total thermal conductivity  shown in Fig.\ref{th}.

\begin{figure}[h]
   \begin{center}
  \includegraphics[width=8.5cm]{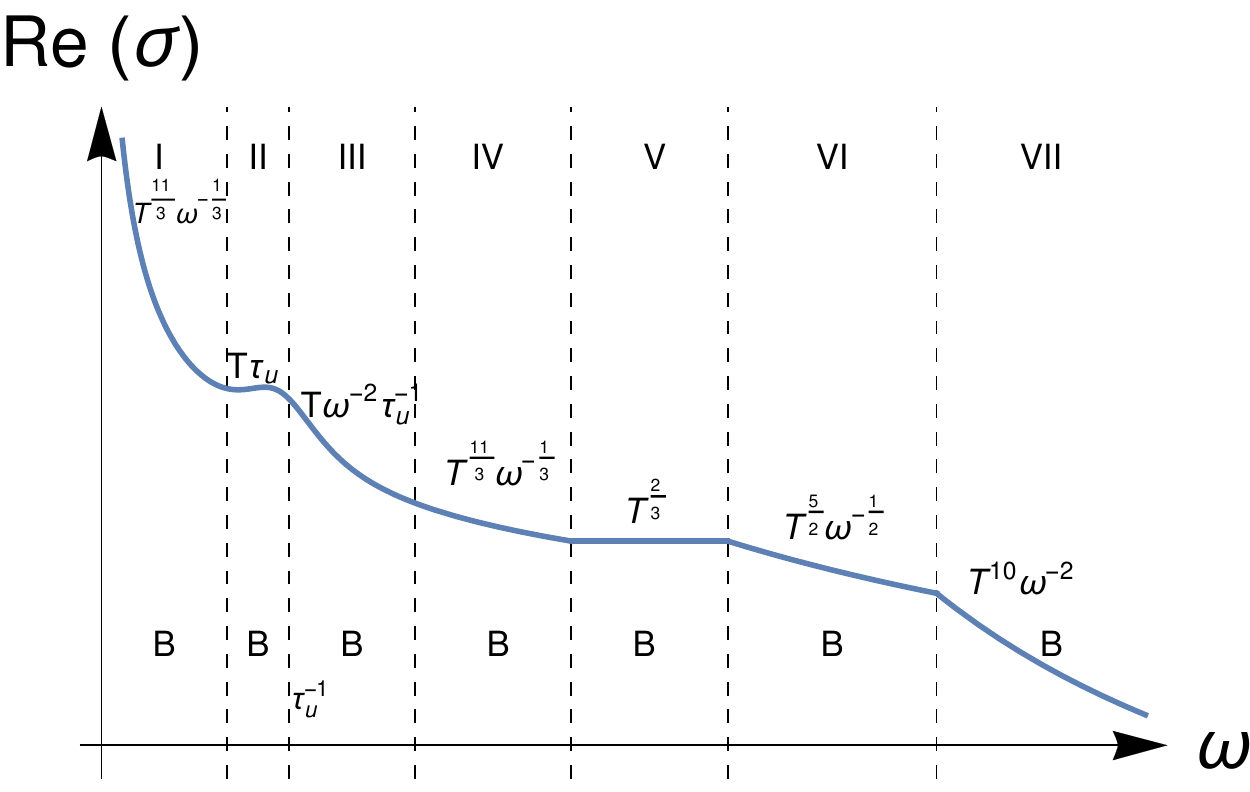}
   \end{center}
    \caption{Real part of $\sigma(\omega)$ at $T_{FB}<T<T_H$, we omit the constants for clarity.
    Label $B$  indicates  that dominant   contribution is bosonic. 
    In Regions II, III  the contribution of the momentum zero mode
 to $\sigma$   is significant, while other regions are dominated by finite energy bosonic modes, e.g. Eq.(\ref{conductivityht}). }
    \label{th}
\end{figure}

For $T \gg T_H$ the bosonic conductivity  $\sigma_B(\omega)$ is given by Eq.(\ref{csht}), thus
\beqa
{\rm Re}~ \sigma^\prime(\omega)=\begin{cases}
              \frac{ T^{\frac{11}{3}} l^{4}  u^{-\frac{5}{3}} m_{*}^{\frac{2}{3}} }{\omega^{\frac{1}{3}}}, \hspace{1.8cm}\omega <\frac{\gamma^3 T^{\frac{17}{3}} l^{-\frac{20}{3}} u^{-16}}{m_{*}^{\frac{34}{3}}},\\\\
                          T^{\frac{16}{9}} m_{*}^{\frac{40}{9}} l^{\frac{56}{9}} u^{\frac{11}{3}} \gamma^{-1}, \hspace{.9cm}{\scriptstyle \frac{\gamma^3 T^{\frac{17}{3}} l^{-\frac{20}{3}} u^{-16}}{m_{*}^{\frac{34}{3}}}<\omega<\frac{\gamma T^{\frac{23}{9}}}{u^{\frac{20}{3}} m_{*}^{\frac{46}{9}} l^{\frac{32}{9}}}},\\\\
          \frac{ T^\frac{14}{5} l^\frac{24}{5} m_{*}^\frac{12}{5} u}{ \omega^{2/5}\gamma^{3/5}               }, \hspace{1.8cm} \frac{\gamma T^{\frac{23}{9}}}{u^{\frac{20}{3}} m_{*}^{\frac{46}{9}} l^{\frac{32}{9}}}<\omega<\frac{\gamma l^7 T^{12}}{m_{*}^4 u^{15}},\\\\
           \frac{ l^2 u^7 m_{*}^4}{ T^2 \gamma             }, \hspace{2.7cm}\frac{\gamma l^7 T^{12}}{m_{*}^4 u^{15}}<\omega<\frac{\gamma l^4 T^9}{m_{*}^4 u^{12}},\\\\
               \frac{ T^{\frac{5}{2}} m_{*}^{2}  u l^4  }{\omega^{\frac{1}{2}}\gamma^{\frac{1}{2}}}, \hspace{2.4cm}\frac{\gamma l^4 T^9}{m_{*}^4 u^{12}} < \omega <\frac{\gamma T^5}{m_{*}^4 u_0^8},\\\\
             \frac{\gamma l^4 T^{10}}{m_{*}^4 u^{11}} \frac{1}{\omega^2}, \hspace{2.4cm}\frac{\gamma T^5}{m_{*}^4 u^8}<\omega. \\
             \end{cases}
 \label{vht}
\eeqa
The resulting  thermal conductivity is schematically shown in Fig. \ref{vth}. 
\begin{figure}
   \begin{center}
  \includegraphics[width=8.5cm]{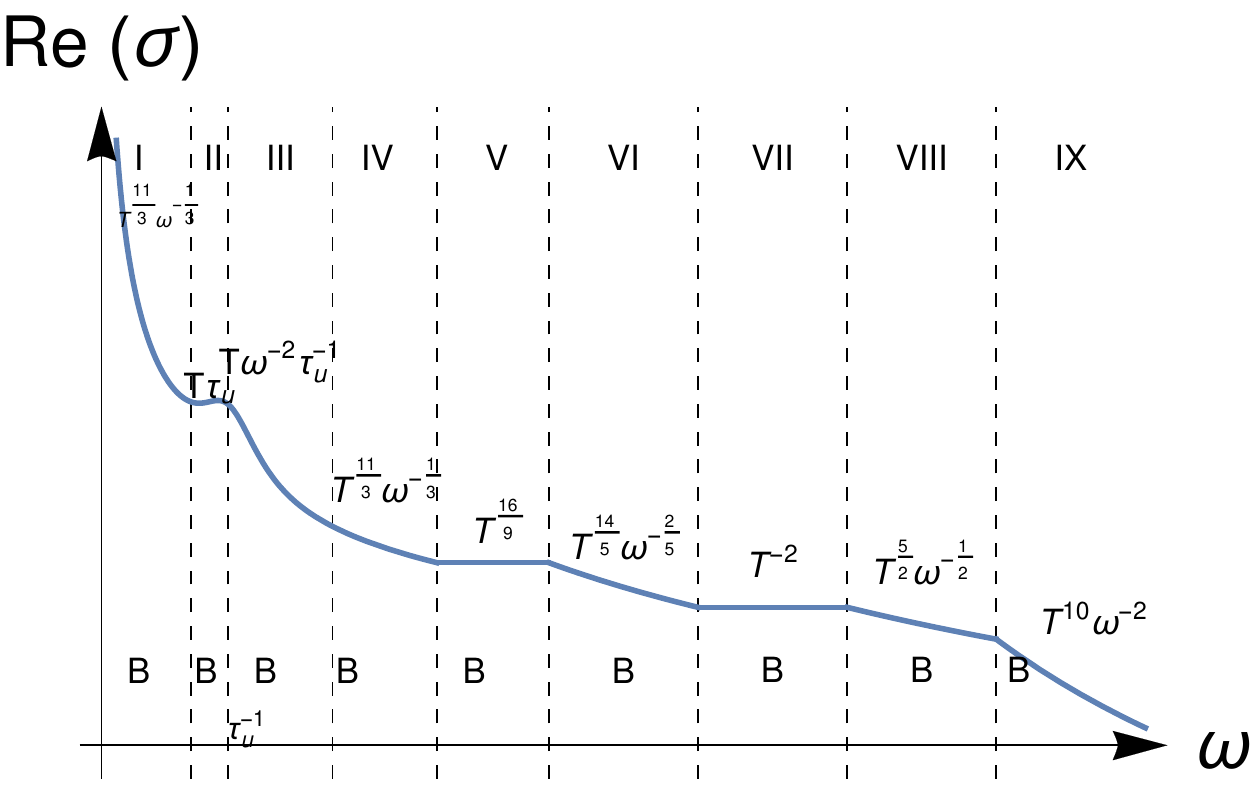}
   \end{center}
    \caption{Real part of $\sigma(\omega)$ at $T \gg T_{H}$, we omit the constants for clarity.
    Label $B$  indicates  that dominant contribution is bosonic. 
   In Regions II, III  the contribution of the momentum zero mode
 to $\sigma$   is significant, while other regions are dominated by finite energy bosonic modes, e.g. Eq.(\ref{vht}). }
    \label{vth}
\end{figure}

 \subsection{Low temperature regime, $T<T_{\rm FB}$ }
 \label{ftclt}
The story is different for $T<T_{\rm FB}$. 
In the  high frequency limit, fermions dominate the thermal conductivity. Below a
critical frequency, $\omega_*=\gamma^3 T^{23}m_{*}^2l^{24}u^{-20}$ bosons dominate it. At the crossover frequency 
$\tau_B(\omega_*) >\tau_F(\omega_*)$, which implies that the bosons are the good quasi-particles and the dominant contribution was found correctly. The inequality remains true even at lower frequencies as the bosonic excitations become more long-lived at low frequencies while the fermionic lifetime remains constant.
The result is 

\beqa
{\rm Re}~  \sigma'(\omega) = \begin{cases}
 \frac{T^{\frac{11}{3}}l^4 u^{-\frac{5}{3}} m_{*}^{\frac{2}{3}}}{\omega^{\frac{1}{3}}},\hspace{1.4cm}\omega<\frac{\gamma^3 T^{23}m_{*}^2l^{24}}{u^{20}},\\ \\

 \frac{u^5}{\gamma T^4l^4}, \hspace{2.4cm}\frac{\gamma ^3 T^{23}m_{*}^2l^{24}}{u^{20}}  < \omega <  \frac{\gamma l^4T^7}{m_{*}^2u^8},\\ \\
   
 \frac{1}{\omega^2} \frac{\gamma T^{10} l^4}{u^{11}m_{*}^4}, \hspace{2cm}\omega > \frac{\gamma l^4T^7}{m_{*}^2u^8}.\\        
\end{cases}
\eeqa
It is worth mentioning that for high  frequencies, the real part of thermal conductivity computed solely with bosonic 
sector matches the one calculated within the fermionic one.
\end{widetext}


\begin{thebibliography}{99}
\bibitem{footnote}The fermionic quasiparticles introduced  in this way are advantageous over the original electrons of the model because their interaction vanishes in the low-energy limit. In particular, they have flat density of states at the Fermi surface.
\end{thebibliography}
\end{document}